\documentclass{article}

\begin{document}
\title{A Multi-proxy Signature Scheme for Partial delegation with Warrant}
\author{Amit K Awasthi and Sunder Lal}
\maketitle
\begin{abstract} In some cases, the original signer may delegate
its signing power to a specified proxy group while ensuring
individual accountability of each participant signer. The proxy
signature scheme that achieves such purpose is called the
multi-proxy signature scheme and the signature generated by the
specified proxy group is called multi-proxy signature for the
original signer. Recently such scheme has been discussed by Lin et
al. Lin's scheme is based on partial delegation by Mambo et al. In
present chapter we introduce a new multi-proxy signature scheme,
which requires less computational overhead in comparison to Lin et
al, and also fulfill the requirement of partial delegation with
warrant simultaneously.
\end{abstract}

\section{Introduction}
Mambo et al introduced a proxy signature scheme in which an original
signer delegates his signing power to another signer called the
proxy signer. They also proposed three proxy signature schemes for
partial delegation based on ElGamal's, Okamoto's and Schnorr's
scheme. Since then several proxy-signature schemes have been
developed
\cite{HCh01:ANMPMSS,HCh01,HSh01:ASMPSS,KPW97,LKK01,LKK01:StrongProxy,MUO96}.
One well-known variation of the proxy signature scheme by Kim et al
includes a warrant in the proxy signature. The warrant is a message,
which includes information of the messages types that are delegated
and proxy signer's identity that prevent the transfer of proxy power
to another party. Another variation, viz, threshold proxy signature
scheme was proposed in \cite{KPW97} in which original signer
delegates its signing power to a set of n proxy signers in such a
way that any $t$ out of these $n$ proxy signers may produce a valid
proxy signature.

\paragraph{}
In some cases, the original signer may delegate its signing power
to the specified proxy group while ensuring individual
accountability of each participant signer. The proxy signature
scheme that achieves such purpose is called the multi-proxy
signature scheme and the signature generated by the specified
proxy group is called multi-proxy signature for the original
signer. Recently such scheme has discussed by Lin et al. In
present chapter we introduce a new multi-proxy signature scheme,
which requires less computational overhead in comparison to Lin et
al, and also fulfills the requirement of partial delegation with
warrant simultaneously. Our scheme is based on Kim et al's scheme.
For ordinary signing operation we use Schnorr's scheme.

\paragraph{}
The rest of the chapter is summarized thus: In section
\ref{RelatedWork}, we give review of literature. In section
\ref{KimsScheme} Kim's scheme is discussed. Section \ref{Proposed}
contains the proposed scheme. Section \ref{proofs} contains proofs
for correct ness of the scheme and in section \ref{performance} we
discuss performance. Finally, Section \ref{conclusion} concludes
the discussion.
\section{Related Previous Works}\label{RelatedWork}
Mambo et al. \cite{MUO96} developed a systematic approach to proxy
or delegated signatures. After the introducing the proxy problem
they firstly divide the delegation in three kinds. which was
further extended by Kim et al \cite{KPW97} by adding one more kind
partial delegation by warrant. Till now we have the following four
types of delegations discussed in previous chapters.
\paragraph{}
Yi et al, in 2000, proposed proxy multisignature scheme which
allows a group of original signers to delegate it's signing power
to a single proxy signer. Proxy signer can sign any message on
behalf of whole group. Further, Hwang et al, in 2001, introduced a
new proxy multi-signature scheme.

\paragraph{}
On the other hand Lee et al's scheme states a multi proxy
signature scheme which fulfills similar criteria discussed as
before. This scheme also allows a group of original signers to
delegate its signing capability to a single proxy signer. We feel
the nomenclature used by Lee et al is not proper. To remove this
ambiguity we define these terms: multi proxy signature and proxy
multi signature as follows:
\begin{quote}
    \textbf{Definition 1}.\emph{An original signer delegates its signing power
    to the specified proxy group while
    ensuring individual accountability of each participant signer. The
    proxy signature that is generated by such a specified proxy group,
    we call, multi proxy signature.}

    \textbf{Definition 2}.\emph{A specified original
    signers group delegate the group's signing power to a single proxy
    signer. The proxy signature that is generated by such a proxy
    signer, we call, proxy multi- signature.}
\end{quote}
\section{Kim's Scheme for Partial Delegation \cite{KPW97}}
\label{KimsScheme} Before introducing our proposed scheme it is
necessary to introduce Kim et al's scheme. Here $p$ denotes a
large prime with $2^{511} < p < 2512$ and $g$ denotes a generator
for $Z_p^*$. Each user selects a secret key $x_u \in Z_q^*$ and
computes a public key $y_u = g^{x_u}~\textrm{mod}~p$.  The system
parameters are -
\begin{itemize}
\item $p$ : A large prime number
\item $q$ : A prime factor of $(p-1)$
\item $g$ : Element of ${Z}_{p}^{*}$ of order $q$
\item $x_u$ : Secret key of the original signer S, here $x_u \in
{Z}_{q}$
\item $y_u$ : Public key of the original signer S, such that $y_u
= {g}^{x_u}$~mod~$p$
\end{itemize}
Basic protocol consists of the following:
\subsubsection{Proxy Generation}
\begin{enumerate}
\item (Key Generation)- An original signer selects a random number
$k \in {Z}_{q}, k \neq 1$ and computes
\[r ={g}^{k}~\textrm{mod}~p~\textrm{and}~ e = h(m_w, r)\]
where $m_w$ is warrant message having the information about
delegation and $h$ is publicly known hash function. Now he/she
computes the proxy signature key $s = x e + k ~\textrm{mod}~q$
\item (Proxy Key Delivery)- The original signer sends $(m_w, s,
r)$ to a proxy signer in a secure way.
\item (Key verification)- After receiving the secret key $(m_w, s,
r)$ the proxy signer computes $e = h(m_w, r)$ and checks the
validity of the key with the following equation \[{g}^{s} = y^e
r~\textrm{mod}~p\] If this congruence passes, he/she accepts it as
secret key otherwise rejects it and request another one, or simply
stops the protocol.
\end{enumerate}
\subsubsection{Proxy Signing}
When the proxy signer signs a message $m$ on behalf of the
original signer, he computes a signature ${s}_{p}$ using any
original signature scheme and $s$ as the secret key. Then the pair
$(m, m_w, s_p, r)$ is the proxy signature,
\subsubsection{Verification}
The verification of the proxy signature is carried out by same
checking operation as in original signature scheme except for
extra computation \[e = h(m_w, r)~\textrm{and}~y' = y^e
r~\textrm{mod}~p\] The value y' may be dealt with as a new public
key, which shows the involvement of Alice.
\section{Proposed Scheme} \label{proposed}
\subsubsection{System Setup}
Choose two large primes $p$ and $q$ such that $q|(p-1)$, a
generator $g \in Z_p^*$ with order $q$. $h(.)$ is a one way hash
function. The original signer O has its private  ket as $x_0 \in
Z_q$ and public $y_0$ where $y_0=g^{x_0}$ mod $p$. $P=\{P_1, P_2,
..., P_n\}$ is the set of the delegated signers. Then each $P_i$
has its private key $x_i$ and public key $y_i$ where $y_i=g^{x_i}$
mod $p$.
\subsubsection{Proxy Generation}
\begin{enumerate}
\item (Key Generation)- The original signer selects $n$ random
numbers $k_i \in {Z}_{q}, k_i \neq 1$ (for $i = 1, 2, ..., n$) and
computes a commitment
\[r_i = {g}^{k_i}~\textrm{mod}~p~\textrm{and}~ e = h(m_w, r)\] where
$m_w$ is warrant message having the information about delegation and
$r = \Pi r_i ~~\textrm{mod}~p$. Now he/she computes the proxy
signature key
\[\sigma = x r+ k ~\textrm{mod}~q \]
\item (Proxy Key Delivery)- The original signer sends $(\sigma,
r)$ to a proxy signer in a secure way and makes $m_w$ public.
\item (Key verification)- After receiving the tuple $(m_w, s, r)$
the proxy signer computes $m_w~=~g^{-\sigma} y_{0}^r r$ mod $p$
and checks the $m_w$. If this holds, he/she accepts it as secret
key otherwise rejects it and request another one, or simply stops
the protocol.
\end{enumerate}
\subsubsection{Proxy Signing}
When the proxy signer signs a message $m$ on behalf of the
original signer, he computes a signature ${s}_{p}$ using any
original signature scheme and $s$ as the secret key. Then the pair
$(m, m_w, s_p, r)$ is the proxy signature,
\subsubsection{Verification}
The verification of the proxy signature is carried out by same
checking operation as in original signature scheme except for
extra computation \[e = h(m_w, r)~\textrm{and}~y' = y^e
r~\textrm{mod}~p\] The value y' may be dealt with as a new public
key, which shows the involvement of Alice.

\section{Performance}\label{performance}Let E, M and I respectively
denote the computational load for exponentiation, multiplication
and inversion. Then following table shows the computational load
of our scheme.
Each phase in our scheme has less computational
load than Lin et al's scheme. In our scheme computational load in
verification phase is 2E + M + H, whereas in Lin et al's scheme it
is 3E + 3M +H. In some applications digital information is signed
once but verified more than once. In such situation the efficiency
of our scheme increases with the number of times verification is
done. Security of scheme is inherited from original scheme used.
Further the total computation cost in our scheme is less than
other existing schemes. Thus, our scheme has computational
advantage over other schemes.
\begin{table}
  \centering
    \caption{Lin et al's scheme}\label{LinTab}

  \begin{tabular}{|c|c|}
\hline
 Phases & Computational Load \\
 \hline
 Proxy generation with   verification & $(n+3)E+(2n)M+H$\\
 \hline
Multi proxy signature generation & $(2n+2)E + (n^2 +2n)M + (n+1)H$ \\
(including proxy verification) & \\
 \hline
Multi-signature verification & 2E + M + H \\
  \hline
 \end{tabular}
\end{table}

\begin{table}
  \centering
    \caption{Our Scheme}\label{OurTab}

  \begin{tabular}{|c|c|}
\hline
 Phases & Computational Load \\
 \hline
 Proxy generation with   verification & $(n+4)E+(2n+4)M+2I$\\
 \hline
Multi proxy signature generation & $(5n+2)E + (4n +4)M + 2H$ \\
(including proxy verification) & \\
 \hline
Multi-signature verification & 3E + 3M + H \\
  \hline
 \end{tabular}
\end{table}

\section{Correctness of the Scheme}\label{proofs}++++++++

\section{Conclusion}\label{conclusion}In present paper we introduced
a new multi-proxy signature scheme, which requires less
computational overhead in comparison to Lin et al, and also
fulfill the requirement of partial delegation with warrant
simultaneously. Our scheme is based on Kim et al's scheme. We also
reviewed the ambiguity in nomenclature of multi proxy signature
and proxy multi signature. and redefine these both signatures on
the basis of literature in existence.
\bibliographystyle{amsplain}
\bibliography{awas,crypto}
\end{document}